\documentclass[manuscript]{acmart}
\usepackage{url}
\usepackage{epigraph}

\setcopyright{none}
\AtBeginDocument{%
  }

\begin{document}

\title{The Commodification of AI Sovereignty: Lessons from the Fight for Sovereign Oil}

\author{Rui-Jie Yew}
\authornote{Both authors contributed equally to this research.}
\email{rui-jie_yew@brown.edu}
\affiliation{%
  \institution{Department of Computer Science, Brown University}
  \city{Providence}
  \state{Rhode Island}
  \country{United States}
}

\author{Kate Elizabeth Creasey}
\authornotemark[1]
\email{kate_creasey@brown.edu}
\affiliation{%
  \institution{Department of History, Brown University}
\city{Providence}
  \state{Rhode Island}
  \country{United States}
}

\author{Taylor Lynn Curtis}
\email{taylor.curtis@mila.quebec}
\affiliation{%
  \institution{Mila --- Quebec AI Institute}
  \city{Montreal}
  \country{Canada}
}

\author{Suresh Venkatasubramanian}
\affiliation{%
  \institution{Department of Computer Science, Brown University}
  \city{Providence}
  \country{United States}}
\email{suresh@brown.edu}

\renewcommand{\shortauthors}{Yew, Creasey et al.}

% \maketitle

\begin{abstract}
 ``Sovereignty'' is increasingly a part of national AI policies and strategies. At the same time that ``sovereignty'' is invoked as a priority for global AI policy, it is also being commodified along the AI stack. Companies now sell ``sovereign'' AI factories, clouds, and language models to  governments, enterprises, and communities — turning a contested value into a commercial commodity. This shift risks allowing private technology providers to define sovereignty on their own terms. By analyzing the history of sovereignty and parallels in global oil production, this paper aims to open avenues to interrogate the implications of this value’s commercialization. The contributions of this paper lie in a disentangling of the facets of sovereignty being appealed to through the AI stack and a case for how analogizing oil and AI can be generative in thinking through what is achieved and what can be achieved through the commodification of AI sovereignty.
\end{abstract}

\begin{teaserfigure}
    \centering
    \includegraphics[width=0.42\linewidth]{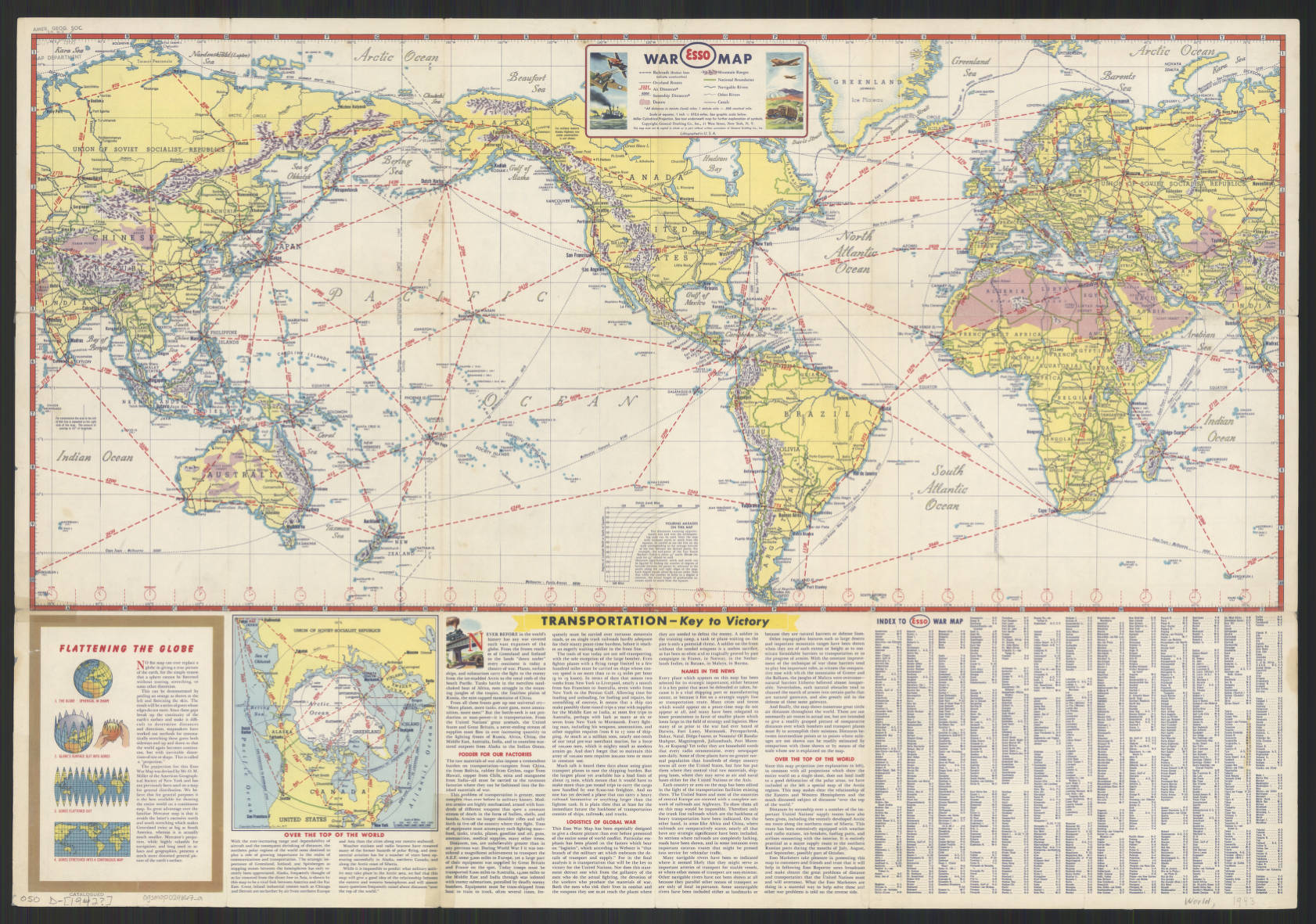}
    \includegraphics[width=0.55\linewidth]{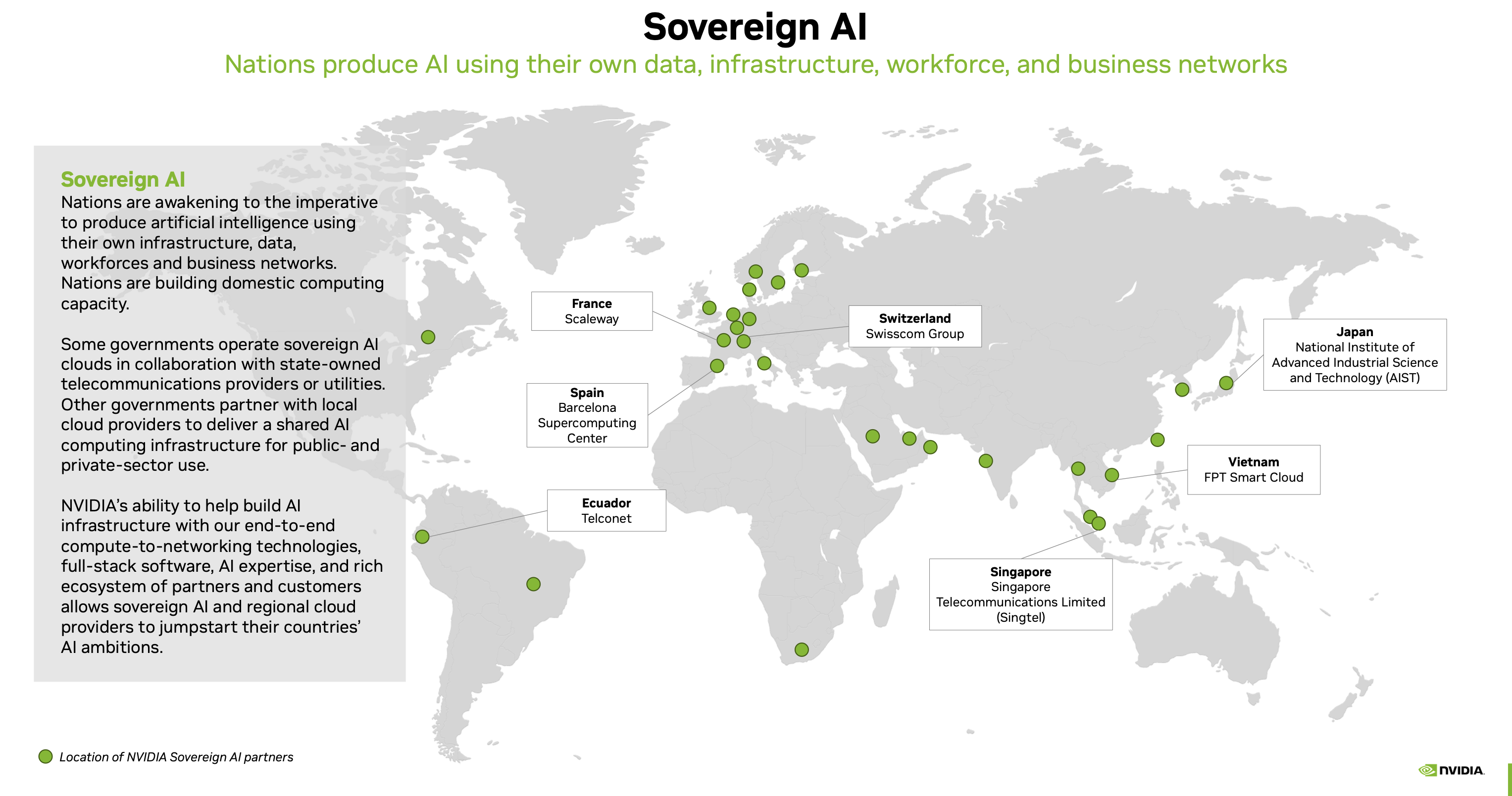}
    \caption{Left: 1942 ``War Map'' produced by Esso, now ExxonMobil, charting oil's role in transportation as  ``key to victory'' ~\citep{EssoWarMap1942}. Right: 2024 map produced by NVIDIA as part of an investor presentation slide deck~\citep{nvidia_investor_oct2024} describing its involvement in sovereign AI efforts globally. On the left panel: ``Nations are awakening to the imperative to produce artificial intelligence using their own infrastructure, data workforces and business networks''}
    \label{fig:maps}
\end{teaserfigure}

\maketitle

\section{Introduction}

\epigraph{\itshape If the 20th century ran on oil and steel, the 21st century runs on compute and the minerals that feed it. This historic declaration hails a new economic security consensus ensuring aligned partners build the AI ecosystem of tomorrow—from energy and critical minerals to high-end manufacturing and models.}{---Jacob Helberg,
Under Secretary for Economic Affairs, \textit{Pax Silica} (December 2025)}

``Sovereignty'' is increasingly a part of AI policies and strategies, from the European Union (EU)'s calls to build a sovereign AI stack~\citep{eurostack_2024}, to India and Singapore’s building of national language model ecosystems~\citep{DST2024BharatGen, singapore_llm_ecosystem_2023}, to Africa’s inclusion of sovereign compute infrastructure as part of its declaration on AI~\citep{africa_ai_declaration_2025}.  At the same time that ``sovereignty'' is invoked as a priority for global AI policy, it is also being commodified along the AI stack. Companies now sell ``sovereign'' AI factories~\citep{nvidia_ai_factory}, clouds~\citep{googlesovereigncloud, flinders2024ibmsovereigncloud}, and language models~\citep{bondarenko2025sovereign, nvidia_sovereign_ai_2025} to national governments, enterprises, and communities --- turning a contested value~\citep{adler2024discursive, jang2025exporting, krasner1999sovereignty} into a commercial commodity. This shift risks allowing private tech providers to define sovereignty on their own terms. 

This paper, in four parts, aims to interrogate the commodification of sovereignty by analyzing the history of the concept and its parallels in global oil production. First, we trace a brief history of sovereignty and how its facets have been operationalized in policy discussion along the AI stack. Second, we motivate the analogical facets between AI and oil we draw from. Third, we map how the commodification of sovereignty along components of the AI stack correspond to notions of sovereignty in history and the potential risks of this discursive and technological strategy. Through comparative historical analysis to the involvement of American and British oil companies in the Middle East in the 20th century, we consider how sovereignty's private technological operationalization along the AI stack risks opening the entities to whom sovereignty is sold to analogous vulnerabilities.

We conclude with recommendations drawn from these parallels --- measuring technological transfer and the tracing of data as a resource. Our goal however is not to sketch out any specific solution or to qualify the value of participating in AI sovereignty initiatives. Rather, we wish to disentangle the discursive facets of sovereignty being appealed to through the AI stack and to present a case for how analogizing oil and AI can be generative in thinking through what is achieved and what can be achieved through the commodification of AI sovereignty. Part of what we are trying to do in this paper is identify ways in which questions about the development and control of the AI stack resonate with older historical processes.
Seen through a historical lens, spaces of criticism become visible that a lens of newness obscures.

\section{A History of Sovereignty and its Role in AI Discourse}\label{sec:aisov}

To understand the relationship between AI and sovereignty, we trace a brief history of the idea of sovereignty and how these ideas have translated into priorities for AI development. Like many major ideas, what sovereignty has meant has changed over time~\citep{ophir2018political}.\footnote{As the political theorist Adi Ophir has observed, concepts are not stable, transcendental objects, rather their meanings change across time and place~\citep{ophir2018political}.} One example of such a definitional shift is from the absolutist monarchies of the 16th century to the ``popular sovereignty'' of the 18th-century revolutions~\citep{Hobbes1651Leviathan, bourke2016popular} --- where the locus of sovereignty shifted from the ruler to the people. 

To disentangle its kaleidoscopic meanings in the AI era, we employ a framework of four distinct ways  sovereignty has been deployed discursively by states: interdependence, domestic, Westphalian, and international legal sovereignty~\citep{krasner1999sovereignty}. These categorizations of sovereignty are borne out of a recognition that states have wielded this concept in their self-interest in many, often conflicting, ways~\citep{krasner1999sovereignty} --- not unlike the ways discourse surrounding AI sovereignty has been characterized~\citep{jang2025exporting, adler2024discursive, lambach2023narratives}. In making sense of how technology companies are wielding this concept, this framework allows us to map how abstract political desires are operationalized into specific AI policy priorities --- such as data localization, sovereign clouds, and multilingual models --- and to consider how these facets are being operationalized technologically. 

\emph{Interdependence sovereignty} is the ability of public authorities to regulate the flow of information, goods, people, or capital across their borders~\citep{krasner1999sovereignty}. The operationalization of this sovereignty focuses on controlling cross-border data flows. This form of sovereignty explicitly hearkens to notions of territorial sovereignty\footnote{Territorial sovereignty has been defined as the exercise of power over a portion of geographic space (containing land, air, and water)''~\citep{storey2017states}}. Just as states control their physical borders, some of the earliest notions of digital sovereignty concerned the control of data flows within and between territories~\citep{hummel2021data}. This meaning traces to the concept of Internet sovereignty'', first invoked in a 2010 Chinese government white paper entitled ``The Internet in China''~\citep{ChinaInternet2010,chander2022sovereignty}. Core to this definition is the assertion that a country should be sovereign over the Internet accessible within its borders and that the data of its residents should be processed domestically~\citep{segal2018china, feng2023chinese}. Current EU sovereignty priorities reflect this same logic, attempting to reassert interdependence sovereignty through initiatives like the EuroStack. These projects aim to support homegrown data storage solutions~\citep{adler2024discursive} and ``sovereign cloud'' infrastructures that keep data resident within the EU~\citep{politico_sovereign_cloud}. Meanwhile, Canada's "Sovereign AI Compute Strategy" explicitly frames domestic infrastructure investment as the only way to "ensure that Canada stays competitive" and independent in the global market~\citep{ised_canada_ai_compute_2025}.\footnote{In another case relevant to data and interdependence sovereignty, the transmission of South Korean user data overseas without requisite guardrails by the Chinese AI company DeepSeek was investigated by South Korea's data protection authority~\citep{reuters2025deepseek}. This incident underscores how lawful data transfers are viewed as requisite to maintaining the state's interdependence sovereignty, its capacity to regulate what enters and exits its digital borders.}

\emph{Domestic sovereignty} is the organization of public authority within a state and the ability of that authority to exercise effective control~\citep{krasner1999sovereignty}. In the context of AI, this control comes in at least two forms: (1) the authority to process and control data and infrastructure relating to territory and identity and (2) economic security and growth.  Notions of data sovereignty have existed in this context since 1998, with Indigenous nations in Canada asserting their rights to sovereignty in the information age against the federal government's claims of authority. These rights were formally asserted through the OCAP\texttrademark{} principles, which outline the inherent rights to Ownership, Control, Access, and Possession of data by Indigenous nations~\citep{FirstNationsCentre2007}. Here, domestic sovereignty is defined not only  by the state's control over the cloud, but by the refusal to cede Indigenous knowledge to federal or corporate entities. Similarly, \citet{floridi2020fight} describes digital sovereignty as a kind of control over digital processes and infrastructure.  Economic security and growth are also appealed to as part of policy priorities, with \citet{lambach2023narratives} noting its prominence as part of German and EU policy narratives.

\emph{Westphalian sovereignty} involves the exclusion of external actors from authority structures within a given territory~\citep{krasner1999sovereignty}. In the context of AI, this manifests as a defensive priority: ensuring that the "mind" of the nation—its data, language, and cultural production—is not dominated by external powers. To understand the stakes of this exclusion, we must overlay the history of European imperial expansion. As anthropologists David Graeber and David Wengrow observe in \emph{The Dawn of Everything}, colonies possessed their own complex histories and local forms of sovereignty long before European arrival~\citep{graeber2021dawn}. From the late 18th to the mid-20th century, European empires worked to systematically undermine these local forms of sovereignty. Imperial agents focused on controlling natural resources and the labor of local populations to serve their own economic interests, actively creating situations of dependency and underdevelopment to cement their control. Concerns surrounding colonialism continue in the digital realm~\citep{coleman2018digital}, with \citet{calzati2022data}, a researcher of media studies and data governance, questioning  whether nations like Kenya actually retain sovereignty as part of digital initiatives that involve entities who may not be driven by the nation's self-interests. 

``Sovereign AI'' priorities have also coalesced around preventing a digital recurrence of this dynamic. One way a kind of assertion of independence, or lack of dependence on foreign technologies, has manifested is through initiatives in Brazil and the EU to build sovereign AI stacks or infrastructure~\citep{eurostack_2024, brazil_ai_plan_2025} and in Africa to compute infrastructure~\citep{africa_ai_declaration_2025}. \citet{filgueiras2025artificial} characterizes the push for AI sovereignty in Latin America as a game between dependence and decolonialization, pointing to the region's infrastructural dependencies.

 The legacy of imperial expansion was not limited to the extraction of physical resources; it also involved the suppression of local knowledge and language systems~\citep{cohn2021colonialism}. In the AI context, there are anxieties that a reliance on U.S.- or China-centric Large Language Models (LLMs) will result in a similar kind of imperialism, where external actors dictate local reality. In South Korea and Taiwan, the push for sovereign AI is motivated by how LLMs developed by dominant firms have responded to politically charged questions~\citep{kim2025sovereign, reuters2025taiwan}.\footnote{OpenAI's GPT-4 stated that the ownership of the Dokdo territory is disputed between Japan and South Korea, a position that conflicts with the South Korean government's stance. This sparked Korean efforts to develop sovereign language models that reflect national historical truths rather than imported equivocations~\citep{kim2025sovereign}.} Beyond political alignment, Westphalian sovereignty also drives efforts to prevent linguistic erasure.\footnote{Models like SEA-LION, developed by the Singaporean government, have been touted for their performance on Indonesian languages, which surpasses that of leading corporate models~\citep{boo2025spore}. By addressing the language challenges that companies and governments face in Southeast Asia~\citep{IBM2024AIsg}, these initiatives assert that the authority to define language and culture must reside within the region, excluding the dominance of Western-centric training data.}

\emph{International legal sovereignty} concerns the mutual recognition of states in the international arena~\citep{krasner1999sovereignty}. In the AI era, this is also manifesting as global AI alliances and collaborations—partnerships that allow nations to pool resources and validate their status as serious technological actors. The quest for technological self-determination often takes the form of building a third pole'' of AI power to compete with and counterbalance the US and China~\citep{chavez2025france}.\footnote{This goal was a central theme of the 2025 Paris AI Action Summit, where French President Macron declared ``technological sovereignty'' a strategic imperative. To achieve this, France has pursued partnerships beyond traditional Western alliances, working with the United Arab Emirates~\citep{France24_2025} and India~\citep{chavez2025france} to build out shared AI infrastructure and capabilities. Similarly, investments like Canada's ``Sovereign AI Compute Strategy'' serve as a signal of national capacity, positioning the state as a recognized peer in the global computing market.} These partnerships extend beyond bilateral agreements to larger coalitions of mutual recognition. The \emph{AI Playbook for Small States}~\citep{AIplaybook}, developed by Singapore and Rwanda under the auspices of the Forum of Small States (a group of 108 countries with populations under 10 million\footnote{This was the threshold for membership when the Forum was initially founded in 1992; some countries have since exceeded that threshold.}), represents a redefinition of sovereignty for smaller nations. Presented at the United Nations Summit for the Future in September 2024, the guide emphasizes that for small states, sovereignty is achieved not through isolation, but through recognized coalitions that can aggregate demand and talent.

\paragraph{Corporate Discourse} Just as \citet{krasner1999sovereignty} points to the multi-faceted aspects of sovereignty  states might deploy in their self-interest, there is similarly a kaleidoscope of meanings being appealed to with ``sovereignty''~\citep{adler2024discursive, lambach2023narratives, jang2025exporting}. AI companies have played a role in constructing the priorities~\citep{jang2025exporting} that surround AI sovereignty and have also  ``learned to offer''~\citep{singh2025rethinking} sovereignty-as-a-service~\citep{grohmann2025sovereignty},  from sovereign clouds to sovereign LLMs ~\citep{singh2025rethinking, yew2025sovereignty}. \citet{jang2025exporting} sounds alarm to how ``sovereign AI'' offerings can serve to belie a global dependency on U.S. firms for AI development while simultaneously allowing for governments and companies alike to signal commitment to independent growth. \citet{grohmann2025sovereignty} argue that Big Tech's offering of sovereignty-as-a-service serves as a kind of discursive co-option of sovereignty. \citet{jiang2024contesting} further note how ``initiatives branded as digital sovereignty may be
frequently used to disguise ambitions to intensify control through digital means''.

In this paper, we similarly analyze the discursive work of sovereignty-as-a-service in the AI context. Our main contributions lie in a mapping of this discourse along different layers of the stack, as well as in the ways we bring forth specific dynamics in oil production and exportation in the Middle East in the 20th century to contextualize the assignment of sovereignty by private AI companies. We also consider how the commodification of AI sovereignty can serve to foreclose some of its more expansive --- or less commodifiable --- meanings.

\section{AI and Oil}
The two primary points of comparison we draw from in this paper are: (1) the tying of oil and AI to economic growth, innovation, and national ambitions and (2) the relational terms of oil production between American and European oil companies in the 20th century and countries in the Middle East\footnote{Specifically, we look at the period prior to the nationalization of oil companies and during which many oil companies had roots in United States and Europe.}  and the relationship between AI sovereignty's commodifiers and those to whom they assign sovereignty. 

 Media materials from companies like Aramco~\citep{elnozahy2022visualizing}, Esso~\citep{EssoWarMap1942}, and Standard Oil highlight the role of oil in modern life  --- emphasizing the ``significance of oil to the emergence of modern civilization''~\citep{graf2018oil}.  Just as oil has been framed as a commodity fueling modernity and  technological innovation, AI, and \emph{sovereign} AI, specifically, is increasingly framed as a source of technological power key to economic growth and national ambitions. It is this promise of modernization and economic growth that also characterizes their production processes as core to conceptions of sovereignty.

 There is also a relative component to the sovereignty a country can have that is captured in discussions surrounding both oil and AI. In discussing the transition to an oil-based energy regime, \citet{dawson2021worker} writes: ``If a particular energy regime and its connected infrastructural assemblage give a country a competitive edge in either
business or military affairs, that energy regime is likely to be quickly adopted and exploited. Nations that do not follow suit are dumped unceremoniously
into the dustbin of history ...  and denigrated ruthlessly as culturally benighted and even degenerate.'' Similarly, in an analysis of the relationship between resources and sovereignty, \citet{stoekl2024sovereignty} describes a notion of sovereignty as malleable and can be dependent on the control of natural resources: `` one nation’s enhanced sovereignty is another’s endangered sovereignty.'' On oil in particular, \citet{graf2018oil} notes that the nationalization of oil in the and the acquisition of sovereignty rights threatened the governments of Western industrialized nations.\footnote{``While the [oil] producing countries [in 1973] acquired rights of sovereignty and coordinated their production policies, the sovereignty enjoyed by the governments of Western industrialized nations appeared to be under threat.''}

Given what has been described as a global ``AI race''~\citep{maizland_fong_2025} and AI's framing as a unit of technological power, as a new form of  electricity~\citep{lynch2017AINE}, as a ``token'' of intelligence~\citep{salvator2025aitokens} --- it is particularly salient that  sovereignty priorities for AI have also coalesced around self-determination and cultural preservation [Section ~\ref{sec:aisov}]. Within  NVIDIA's  broader vision of the building of ``AI factories'' as an analogue to electricity factories and ``industrial infrastructure''~\citep{janakiram2025ai}, sovereign AI is embedded into a broader story that, just as electricity factories spin water into electricity, AI factories--comprising of NVIDIA chips--spin data and energy into AI, which then translates into important economic and geopolitical advantages, or into preventing the loss of important economic and geopolitical advantages~\citep{nvidia_ai_factory}. As regards to our second point of comparison, as part of concession agreements, in the 20th Century, oil-exporting companies contracted for exclusive land rights in oil-producing territories. The control that oil companies had over territories through concession agreements challenged notions of territorial sovereignty.  Territorial sovereignty has traditionally been considered the primary realm of a state's exertion of sovereignty.\footnote{As part of a ``classic definition of a state''~\citep{eichensehr2018digital} presented in~\citep{gerth1946politics}, a state ``claims the monopoly of the legitimate use of physical force within a given territory''. And, while companies may be headquartered or incorporated in a territory, they ``do not possess sovereign territory''~\citep{eichensehr2018digital}.} In the case of AI, companies are contracting with  governments and other entities in the production of ``sovereign AI''. We speak to the way in which commodifiers of AI sovereignty are asserting technological control that analogously challenge notions of sovereignty that have rested in the state. To further illustrate the basis of our comparison, we include two maps in Figure~\ref{fig:maps}, one produced by Esso, Standard Oil of New Jersey, now ExxonMobil, in 1942, the other produced by NVIDIA in 2024. Both maps highlight the global implications of its respective commodity and the importance of that commodity to meanings of sovereignty.

 The analogy between oil and AI has a longer history, however, with the relationship between data, a crucial component of AI systems, and oil extending as far back as 2006.\footnote{``data is the new oil'' is said to have originated in 2006 from Clive Humby~\citep{ICODataCommodity}. \citet{hirsch2013glass} has also pointed to the externality similarities between oil spills and data breaches.} The turn to the 2020s marked a shift in understanding AI production as requiring the extraction of physical commodities~\citep{crawford2021atlas,  hao2025empire} --- not unlike the production of oil~\citep{khalili2025extractive}.\footnote{Indeed, as recently as 2011, the impact of Big Tech was conceptualized as almost separate from the physical realm.  \citet{chander2011facebookistan} writes, 
``Facebook’s physical manifestation
in people’s lives is through LED screens, not soil.'' 2011, when the above article was published, was also the year Facebook opened its first data center facility in Prineville, Oregon~\citep{metz2011welcome}, which was chosen for characteristics like the cool and dry quality of its air.} It is now apparent that the current paradigm of AI production cannot be divorced from its effects on the physical realm: it is dependent on the large-scale extraction of physical resources~\citep{crawford2021atlas} and on data centers~\citep{sayegh2024billion}, whose ``physical manifestation'' has since had considerable impact~\citep{copley2025data}.
While important, our paper's focus is not on the increasingly overlapping material and energy concerns between AI and oil, but on the analogies we discuss above.

\paragraph{Limits of Our Analogical Lens}
The analogy between the oil and AI contexts we describe is not perfect. One major difference is that foreign oil companies contracted with governments in order to gain access to the source of the commodity to be refined --- oil. Embodied within these terms is a relational structure that placed host territories as \emph{producing} a resource that foreign companies were explicitly purchasing --- and, as some scholars have argued --- extracting~\citep{khalili2025extractive}.  On the other hand, as part of sovereign AI offerings and discourse, AI companies are positioning themselves as \emph{offering} and developing this resource. This makes the relational dynamic between foreign companies and sovereign entities appear different in these two contexts of oil and AI. The production of oil as a physical commodity is tied to where it can be found, differentiating it from AI, which does have material implications~\citep{crawford2021atlas} but is framed as something that can be ``grown'' in every country~\citep{caulfield2024nvidia}. However, the positioning of AI companies as offering a resource can also serve to obscure the mechanisms through which AI companies can instead extract resources through these ventures from those to whom they discursively assign sovereignty. Part of our intervention in this piece is to consider how this arrangement opens up entities to whom ``sovereignty'' is granted to analogous vulnerabilities in the oil context.

\section{``Sovereignty'' Along the AI Stack}\label{sec:sovstack}
The technological operationalization of AI sovereignty tends to occur around an AI ``stack''~\citep{belli2026ai}: the assemblage of software and hardware components that together make up a top-to-bottom AI system implementation. \textbf{Through parallels in the myth-making and operation of foreign oil companies operating in the Middle East in the 20th century, we map out facets of how this private discursive construction of a sovereign AI stack may serve to distort sovereignty into a value that can be addressed on technological terms that serve to expand the reach of the private sector.} In alignment with  \citet{jang2025exporting}'s findings that NVIDIA is one of the earliest and primary drivers of ``sovereign AI'' discourse, our analysis focuses in no small part on NVIDIA's ``sovereign AI'' initiatives. To conduct our analysis, we use the search engine in NVIDIA's corporate blog, analyzing blogposts with the ``sovereign'' tag from 2019 to December 2025. We also analyze materials from Google, Microsoft, and Amazon, three of the biggest cloud providers comprising ``Big Cloud''~\citep{Widder2025HowBigCloud} in order to lay out their sovereign cloud offerings, and materials from Meta, Cohere, and HuggingFace, major open-source and multilingual large language model developers, to analyze multilinguality and open-source and how these offerings are tied to sovereignty goals. 

\subsection{Hardware and Infrastructure}\label{sec:hardware}

Chips and the raw materials they are comprised of form the basis of the AI stack, enabling the computational processing to train AI and to power the technology across its stages of development and deployment.\footnote{At the highest level, central processing units (CPUs) are general-purpose chips that support a wide range of languages and applications~\citep{zhou2024study}. However, CPUs encounter significant slow-downs when processing the massive amounts of data and computations necessary to produce AI.  Application-specific integrated circuits (ASICs) are designed for one specific purpose and are also vital to industrial automation systems~\citep{smasic}. Graphical processing units (GPUs), which are a type of application-specific integrated circuit (ASIC) chip~\citep{mchugh2024techspert}, are chips specialized to enable the parallelization of calculations required in AI training~\citep{khan2020ai}. 
Other chips relevant to AI production include tensor processing units (TPUs), also a kind of ASIC chip, which can run highly specialized to run AI-based compute tasks~\citep{mchugh2024techspert}. Field-programmable gate arrays (FPGAs), which can re-programmed by downstream developers to handle different customized use cases, have been used to power search applications ~\citep{mcmillan2014microsoft}.} NVIDIA, the dominant player in the AI chips market and a company headquartered in the United States~\citep{levitt2024nvidia}, has responded to and pushed regulatory priorities and geopolitical concerns surrounding AI by announcing their commitments to ``sovereign AI'', which the company refers to as ``a nation's capabilities to produce artificial intelligence using its own infrastructure, data, workforce, and business networks''~\citep{lee2025sovereign}. NVIDIA CEO Jensen Huang has stated that ``every country needs sovereign AI''~\citep{caulfield2024nvidia} and has highlighted the need for every country to ``own the production of their own intelligence.'' The company has also pursued this story through establishing conferences in the name of sovereign AI, like the Sovereign AI Summit at the company's annual GTC Conference, which explored ``how nations are embracing AI factories as critical infrastructure for digital transformation and resilience''~\citep{nvidia2025gtcsovereign}.
The discourse is widely deployed as part of NVIDIA's global governmental initiatives, through which it is working with over ten countries in the development of AI infrastructure as part of their sovereign AI initiatives --- including India, Denmark, Thailand, New Zealand, Switzerland, and Japan~\citep{yang2024jensen}. This thread of discourse continues around initiatives by Grok and OpenAI to build up national AI infrastructures ~\citep{openai2025countries, openai2025stargate, xai2025grokglobal}. In 2025, OpenAI announced its ``OpenAI for Countries Initiative''~\citep{openai2025countries}, through which the company is laying down compute infrastructure for Stargate initiatives in Norway and the United Arab Emirates (UAE). In addition to appealing to the technological competitiveness of Norway and the broader European region\footnote{As described in the Stargate Norway announcement, OpenAI and its Stargate Norway partners ``will also work to provide priority access to Norway’s AI ecosystem, ensuring homegrown AI start-ups and scientific researchers can benefit from the additional compute capacity. Surplus capacity will be made available to public and private sector users across the UK, Nordics and Northern Europe, serving regional demand and accelerating the development of Europe’s AI ecosystem''~\citep{openai2025stargate}}, the Stargate Norway initiative announcement also specifically appeals to ``sovereign AI'' and the  `` benefit of [Norway's] people''~\citep{openai2025stargate}, hearkening to notions of popular sovereignty and appealing to notions of Westphalian and international legal sovereignty --- appealing to the recognition of the these countries' technological authority.

Impression management through media materials and corporate social responsibility initiatives similarly enabled Aramco, at the time an American company, to navigate its oil business in Saudi Arabia in the 20th century~\citep{albalwi2024pro, elnozahy2022visualizing}. Learning from Standard Oil's brush with antitrust enforcement in the United States~\citep{elnozahy2022visualizing}, Aramco took seriously its public relations arm,  hiring historians and political scientists~\citep{barrett2014aramco} to construct appealing stories around its activities in the Middle East and honing in on its role in a kind of ``corporate enlightenment''~\citep{barrett2014aramco}. \textbf{Like the discourse surrounding NVIDIA's sovereign AI initiatives, these stories centered  similar themes of partnership, and the role of science and technology in national modernization.} These themes correspond with understandings of domestic and international sovereignty and its discussion in relation to resources in ensuring that nations have the resources and wherewithal to exercise authority~\citep{krasner1999sovereignty}.

Just as NVIDIA has leaned into the language of  ``partnership'' as part of its sovereign AI pursuits~\citep{NVIDIA_EuropeAI_2025, Hacker2025DeutscheTelekom, Martin2025Korea}, Aramco also adopted the language of partnership, calling its endeavors the ``Arabian- American Partnership''. The language of ``partnership'' with Saudi Arabia, is further exemplified in the company's change in name from Standard Oil of California to Arabian American Oil Company (Aramco). The language of partnership enabled Aramco to project an image as a partner in Saudi Arabian nation-building, positioning the company for a strong relationship with the country's monarchy --- ``whose continued backing Aramco required if
it hoped to exploit the kingdom for its riches''~\citep{barrett2014aramco}.

On its corporate blog, NVIDIA displays how its sovereign AI initiatives are catapulting countries to AI ``leadership''~\citep{hills2025ukai, nvidia2025saudi}, as well as how the company is supporting the ``global policy imperative to harness domestic AI capabilities that are critical to economic growth, national security, cultural preservation, and innovation''~\citep{nvidia_global_public_sector}. Aramco similarly leaned into the rhetoric of science and technological wherewithal as a way for the Saudi Arabian government to expand its authority and fulfill its national ambitions: ``[Aramco] worked alongside Saudi Arabia's king 'Abd al-Aziz ibn Sa'ud as he turned to science and technology to expand his authority. It was a nation-building dependent on new technologies for communication, transportation, and administration, and Aramco situated itself as an indispensible partner in this social and economic development''~\citep{barrett2014oil}. For Aramco, this myth-making enabled both the company putting forth this rhetoric and the country's monarchy to benefit in their claims of territorial sovereignty.\footnote{``The monarchy, in turn, encouraged this narrative
as it sought to consolidate its rule and assert ever- expanding land claims''~\citep{barrett2014aramco, jones2010desert}. ``Narrative serves as a powerful tool to a corporate entity interested in expanding
its presence along the frontiers of the Rub al- Khali, or Empty Quarter, as the Americans called the southern Arabian desert, and narrating the new Saudi state into the
distant past. By re-imagining the historical connections between the land and the
people living on it, Aramco's historians were able to recast widely held notions of
territorial claims and compel the great powers to grapple with previously `settled'
issues.''~\citep{barrett2014oil}} The story of sovereign AI put forth by NVIDIA can similarly enable kind of co-construction that can simultaneously be wielded by governments and to influence them to open up territorial resources.\footnote{In particular, \citet{jang2025exporting} notes that the use of ``sovereign AI'' benefits governments by enabling them to signal commitments to technological development.}

The promise of ``sovereign'' AI through the development of ``sovereign AI factories''~\citep{nvidia_ai_factory}  seems a straightforward transaction: companies provide the computational infrastructure for ``sovereign AI''; the countries open up their land and resources to support it. This often requires upping energy capacity and electricity generation. Leaning into the imperative of sovereign AI,  NVIDIA has lobbied the Japanese government to generate more power to fuel Japan's AI infrastructure~\citep{kiyohara2025nvidia}, with the company's CEO saying ``[t]he country needs to build new infrastructure'' and pointing to the energy requirements of ``generating and creating intelligence''~\citep{kiyohara2025nvidia} --- just as Aramco's oil story enabled the company to gain access to territorial resources. At the same time, hardware and infrastructure providers are often not just in the business of laying down computational infrastructure. They are also in the business of training and deploying models ---and \emph{\textbf{training countries and enterprises to develop and work with their hardware and models}}. As part of sovereign AI initiatives in India, NVIDIA has already trained over 100,000 AI developers and created over 2,000 ``NVIDIA inception AI companies''~\citep{takahashi2024nvidia}, wherein these companies use models like Nemotron-4-Mini-Hindi-4B to provide their services~\citep{takahashi2024nvidia}. 

In Saudi Arabia, as part of a sovereign AI initiative including the Saudi Arabia Data Authority, NVIDIA is additionally ``train[ing] government and university scientists and engineers on how to develop and deploy models for physical and agentic AI''~\citep{nvidia2025saudi}. OpenAI and Grok are also launching educational initiatives surrounding their global AI offerings. ChatGPT Edu in Estonia, for example, was announced in early 2025~\citep{openai2025estonia} and ``Grok Goes Global'''s educational program with El Salvador's government and its public schools was announced in late 2025~\citep{xai2025salvador}.\textbf{These companies are then supplying not only AI's infrastructural basis, but also models atop this infrastructure, and the education surrounding these components. A story doused in technical language and the promise of technological leadership~\footnote{``Saudi Arabia to Become a Global AI Leader With Sovereign AI Infrastructure Powered by NVIDIA''~\citep{nvidia2025saudi}} can serve to bury a reality in which they may not be sovereign over how and the degree to which their territory and resources are used to produce AI and to what ends.}

\subsection{The Cloud}\label{sec:cloud}

Three cloud companies --- Amazon Web Services (AWS), Microsoft Azure (MA), Google Cloud (GC) --- control two thirds of global cloud compute market~\citep{Widder2025HowBigCloud}, with \citet{Widder2025HowBigCloud} terming this collection of companies ``Big Cloud''.  AWS and MA each offer of a sovereign cloud for the EU~\citep{amazon2025awseu, brill2025microsoft}, and GC has released a blogpost detailing the ways in which EU customers are benefiting from sovereign cloud services~\citep{brady2024eu}. Recurring behind each of these initiatives is a rhetoric that the EU sovereign cloud they offer will not only help the EU maintain control over data on the cloud by keeping EU data within EU borders and by ensuring that data transfers across its borders occur securely\footnote{E.g., ``[w]ith the completion of the boundary, our European commercial and public sector customers are now able to store and process their customer data and pseudonymized personal data for Microsoft core cloud services — including Microsoft 365, Dynamics 365, Power Platform, and most Azure services - within the EU and EFTA regions''~\citep{brill2025microsoft} and ``[t]he AWS European Sovereign Cloud will allow customers to keep all customer data and the metadata they create (such as the roles, permissions, resource labels, and configurations they use to run AWS) in the EU''~\citep{amazon2025awseu}.}, but will also bolster the EU's economic competitiveness amidst a technology industry dominated by American firms.\footnote{E.g., `` [i]n total, AWS’s planned investment is estimated to contribute {17.2} billion to Germany’s total Gross Domestic Product (GDP) through 2040, and support an average 2,800 full-time equivalent jobs in local German businesses each year''~\citep{amazon2025awseu} and ``[w]e will help strengthen Europe's economic competitiveness, including for open source''~\citep{microsoft2025europeandigital}.}.  The rhetoric of control over data within a territory's borders appeals to notions of domestic and interdependence sovereignty --- appealing to the EU's ability to control data flows and within their borders ----  and the rhetoric of economic competitiveness appeals to Westphalian and international legal sovereignty --- appealing to the recognition of the EU's authority. Moreover, sovereignty as conceptualized by these CSPs does not only implicate governments --- it is also sold to other kinds of enterprises and organizations.\footnote{\citet{gcp2024foundations} offers sovereignty to all: ``We built carefully designed controls, software and hardware controls, to ensure that organizations across all sectors can benefit from the magic of the best cloud, data, and AI technologies available without having to give up on their strategic autonomy ambitions.'' In an IBM blogpost, \citet{flinders2024ibmsovereigncloud} lays out benefits of a sovereign cloud for other enterprises, including limiting access to customer data for employees. In fact, ``self-sovereign AI'' as supported by NVIDIA's cloud services is also described to appeal to enterprises, providing businesses an avenue to ``commercialize their work'' and to provide AI services that are ``secure, transparent, and verifiable'' --- which the company also describes as ``important in the face of governmental scrutiny''~\citep{nertney2024nvidia}.}

As part of its business model, CSPs rent out access to computation provided by their physical data centers through subscription models~\citep{cobbe2023understanding}, resulting in a loss of control from the exporting of data to the cloud.\footnote{For instance, \citet{wang2019riverbed} note that, when data is ported to the cloud, ``the user loses control over where data is stored, how it is computed upon, and how the data
(and its derivatives) are shared with other services.''} Privacy and security have been important aspects of providing that control to users or those whose data is ported to the cloud, and privacy-enhancing technologies (PETs) have been framed as addressing issues relating to this lack of control. As \citet{gurses2017privacy} predicted, CSPs have now begun to treat privacy as another dimension on which to compete, with CSPs touting offerings across a range of different PETs~\citep{aws_wickr, azure_differential_privacy}. But control is not only a dimension that has been important to notions of digital privacy. It is also 
a trigger of liability across different regulations --- and CSPs have strategically wielded and relinquished control to manage that liability.\footnote{\citet{cobbe2024governance} notes that the terms of service of CSPs are arranged to ``decouple legal rights
and benefits around operation and control of technologies from legal risks and
responsibilities, arranging them to instead maximise providers' commercial benefit and
externalise liabilities and accountabilities to the limited control delegated to customers.''}  This dynamic has played out in practice. A 2024 complaint by the non-profit group noyb against Microsoft \citep{lomas_microsoft_2024} alleges that Microsoft evaded its responsibility of fulfilling data transparency and access requests. Leveraging  relational structures in data protection and privacy laws, the company claimed that that school was the controller of its data and that it was merely a processor of the school's data --- even when the local school did not have control or visibility into Microsoft's operations to fulfill those requests. Scholarship has also explored how encryption, traditionally understood as a security technology, has been used to manage knowledge, and, thus, liability~\citep{veale2023denied}, sometimes even under the pretexts of privacy or security protection~\citep{van2022privacy}. 

Technologies that have historically been treated as tools for privacy and security by CSPs are being cast and even re-cast as a ``sovereignty controls''.\footnote{GCP lists encryption as a part of their sovereignty offerings~\citep{gcp2024foundations}. NVIDIA has also listed properties like confidentiality and decentralized computing as a property of ``self-sovereign AI''~\citep{nertney2024nvidia}.}  
At a time when privacy and security technologies are facing increased governmental scrutiny\footnote{Governments like the United Kingdom~\citep{kubi_encryption_2025} and  Sweden~\citep{nojeim_lorenzo_perez_2025} have requested that companies develop an encryption ``backdoor''~\citep{wu_chung_yamat_richman}, which would then enable these governments to access the data these companies possess.}, the framing of privacy and security technologies as sovereignty-enhancing can tap into the connotations of sovereignty as government-serving to elide the undertones that make security technologies unsavory to regulatory entities~\citep{vallance2023signal, diffie2010privacy}.\footnote{Even end-to-end encryption, a technology that has sparked controversy in governmental circles, has been painted as enabling ``clients to meet their sovereignty needs''~\citep{google_workspace_2025}.  We also borrow language from~\citep{gillespie2010politics}, which describes similar behavior in a different context.}. In appealing to all organizations, CSPs' assignments of sovereignty to one entity --- e.g., the government --- can clash with the assignment of sovereignty to another --- e.g., corporations. \textbf{The tensions inherent in these simultaneous invocations are simply swept away. Indeed, sovereign AI developers can, at once, appeal to the discourse of sovereignty as a way to give governments control over AI development, while selling to corporations a way to guard against government control over their AI development.\footnote{NVIDIA sells  ``sovereign AI'' to governments around the world~\citep{nvidia2025saudi}, while also selling ``self-sovereign AI'' that it highlights to corporations as a way to handle governmental scrutiny~\citep{nertney2024nvidia}.}} Moreover, under the backdrop of policy developments that afford the U.S. government increased scrutiny over data transmissions~\citep{cloudact2018}, \textbf{``sovereignty controls'' embedded within compute infrastructure can allow these CSPs, mostly based in the U.S., to play both sides: to give external enterprises and governments the sense that their data remains secure from external access even as that data remains vulnerable to U.S. visibility and control~\citep{TheRegister2025Microsoft, SenatFrance2025CE}.}  

A parallel use of technology --- mathematical concessionary formulas --- enabled the Anglo-Persian Oil Company (AIOC) to elide political tensions and retain control over oil profits in Iran. A central avenue through which profits flowed through the production and transportation of oil was through concessionary agreements. The ownership of oil pipelines in the Middle East can be traced to colonial powers~\citep{havrelock2017borders}. Like how American cloud companies, pipelines of data, are re-branding their corporate initiatives to bear the European name~\citep{microsoft2025europeandigital, googlesovereigncloud, microsoftcloudforsovereignty, amazon2025awseu}\footnote{At a 2025 cloud summit, the French digital minister herself called out the way that American companies were partnering with the EU as a form of ``sovereignty-washing''~\citep{theophane2025against},. Indeed, there is increasing recognition that these initiatives can constitute a kind of  ``sovereignty washing'', where American firms are claiming that the EU has control over the sovereign cloud but where the firms themselves have control over cloud operations in actuality~\citep{adler2024discursive}.}, American and European oil companies operating in the Middle East did the same. On the Iraq Petroleum Company's operations in Iraq after World War I, \citet{havrelock2017borders} writes that ``[t]he only thing Iraqi about the company [that] was the location of the oil'': the Anglo-Persian Oil Company (renamed the AIOC and, now, BP) and Shell, both British companies, held a collective $43.75\%$ of the profits surrounding the oil being transported through the pipelines. \citet{shafiee2012petro} notes that the first half of the 20th-Century was marked by dramatic crises and relational tensions between the AIOC and the Iranian government. Concessionary disputes ``erupted around production labor and profits''~\citep{Shafiee2022MachineriesSeminar} and these crises were responded to through the ``technology of a formula'', including formulas which calculated royalties to the Iranian government,  which allowed APOC to ``negotiate possible compromises'' ~\citep{shafiee2012petro},  to construct terms of negotiation and foreclose certain political possibilities.\footnote{``The company responded oftentimes by constructing mathematical formulas but also legal arguments and scientific representations about the oil to create these new spaces for negotiation and to manage political outcomes more favorably to support its concessionary authority ...''~\citep{Shafiee2022MachineriesSeminar}}

The complexity and technicality of these formulas also served to manage and elide tensions within the interests of multiple stakeholders. As \citet{shafiee2012petro} writes, these formulas ``served as calculative technologies equipping AIOC
officials and accountants with the capacity to redirect political outcomes
against competing socio-technical programmes put forward by Iranian oil
workers, the communist party, reformist nationalism and public opinion
demanding more democratic forms of oil production.'' Importantly, even as contracts negotiating the relationships between oil-producing and oil-exporting nations evolved to apportion more authority to oil-producing nations, these formulas served to foreclose lines of questioning that could threaten British control over oil reserves~\citep{shafiee2012petro}. In the 1970's, new contract forms evolved to replace oil concession agreements~\citep{zorn1983permanent}. However, even as the formal terms of ownership and control appeared to favor oil-producing countries, the countries remained limited by how much access they had into and ability to scrutinize company operations: ``[w]ithout such capacity, countries may find it impossible to exercise more than token authority over the oil companies.''

Finally, despite CSPs' offering of sovereignty to its clients, CSPs are wielding control --- arguably, even sovereignty --- through the exercise of its capacity to cut off access to client services~\citep{biddle2025microsoft} as well as in their discretion surrounding responsibilities to turn over data to the U.S. government under regulations like the CLOUD Act ~\citep{woollacott2025microsoft}. The authority to turn off access speaks to the exercise of an understanding of sovereignty defined by political theorist Carl Schmitt as the right to decide on the exception, or when the rules do or do not apply~\citep{daston2022rules}.\footnote{ ``German political theorist Carl Schmitt (1888 - 1985) famously defined sovereignty as the power to decide on the exception. Schmitt was adamant that the exception, which arises in times of ‘extreme peril’, cannot ‘be codified in the existing legal order.’ Unlike casuistry, which tests one rule against another, or equity, which bends the letter of the law to conform to its spirit, which bends the letter of the law to conform to its spirit, the state of exception proclaimed by the sovereign breaks with rules altogether''~\citep{daston2022rules}.
}

\subsection{Model Building and Application}\label{sec:model}

At this stage of the AI stack, the discourse of AI sovereignty manifests through sovereign large language models (LLM), particularly LLMs that can process or generate multiple languages (multilingual models)\footnote{For example, NVIDIA labels its fine-tuned multilingual models as ``sovereign-ai''~\citep{nvidia_sovereign_ai_2025}.} or languages that are under-resourced. This includes the capacity to process text on that language, but it also encompasses the capacity to generate or respond to text in a way that captures different cultural and political nuances.  Rhetoric surrounding the private development of LLMs connotes values like cultural preservation that are core to understandings of indigenous sovereignty~\citep{iorns1992indigenous, smart2024socially} and aspects of international legal sovereignty in emphasizing a language's external status or asserting the recognition of a political perspective.   

As regards to every country's need for sovereign AI, NVIDIA highlights the role of AI sovereign in ``codif[ying] your culture, your society's intelligence, your common sense, your history''~\citep{caulfield2024nvidia}. ``Sovereign AI'' has also been painted as an effort in providing services in more languages and with an understanding of cultural nuances.\footnote{NVIDIA's offerings surrounding fine-tuning language models for translation quality~\citep{du2025nvidia} and text data processing for under-resourced languages, like those in the Southeast Asian region~\citep{bleiweiss2025mastering} have been painted as ``sovereign'' AI or ``sovereign'' LLMs. Increased performance and support for under-resourced languages like Vietnamese are then discussed as  ``ensuring a seamless and impactful customer experience'' for enterprises~\citep{bleiweiss2025mastering}.} Details surrounding Meta's language model and dataset releases are released within a research paper titled ``no language left behind''~\citep{costa2022no}.\footnote{\citet{costa2022no} additionally includes interviews with subjects who are members of communities whose languages are under-resourced and highlights the need to ``prioritize the need of under-served communities'' and motivates the benefits of the incorporation of low-resource languages machine translation as  ``help[ing] de-prioritized languages gain digital
visibility on a global scale, which could compel local institutions to take native languages
more seriously and invest more resources into preserving or teaching them''~\citep{costa2022no}.} Similarly, a policy primer from Cohere has pushed the need for multi-lingual models so that low-resourced languages do not get left behind~\citep{cohere2024ai_language_gap}. And, as part of a paper led by Google researchers on ``socially responsible data for large multilingual language models'',  \citet{smart2024socially} also emphasize the role of LLMs in preserving under-resourced languages and integrating them within a broader language network: ``As NLP and LLMs advance, there is an opportunity to assist communities not only with the preservation of their languages but also with empowering integration and engagement of their languages with globally connected networks --- in sharp contrast to how many have been historically excluded from major global systems or relegated to their margins.''

The imperative of sovereign AI under the development of multilingual models~\citep{nvidia_sovereign_ai_2025} and the header that ``no language should be left behind''~\citep{costa2022no, cohere2024ai_language_gap}, can serve to uphold an extractive model of data labor --- namely through the use of data annotators --- under the guise that the resulting models will preserve cultural values and benefit these communities. 
Data annotators are a critical component of AI production and the sourcing of  ~\citep{gray2019ghost}. Annotators are typically sought through third-party platforms and in developing nations, and they perform tasks ranging from ranking translations in languages they are fluent in~\citep{deck2023scale} to identifying and labeling objects in images~\citep{miceli2020between}. In addition to pay discrepancies along the lines of language --- with Scale paying its annotators for German $\$21.55$ an hour and its annotators for Telugu $\$1.43$ an hour~\citep{deck2023scale} --- the consequences of this labor can be severe. The impacts of AI on translator roles are especially felt in Turkey, where translators have traditionally played a respected role in the country's diplomatic history~\citep{genc2025translators}. 
Data annotators have also reported traumatic disorders as a result of the nature of their work~\citep{newton2023ai}. And, because this labor is task-based and often goes unacknowledged~\citep{gray2019ghost}, there is also often a degree of financial and labor precarity involved, with Scale AI, a data annotation platform used by developers of multi-lingual models like Google, Meta~\citep{techcrunch_google_scale_2025}, and Cohere~\citep{scale_cohere_2024}, revoking the access of entire countries to their annotation platform with no explanation overnight~\citep{restofworld2024scale}.

Oil production also created specific forms of labor precarity and repression. Looking at the context of BP’s operations in the Persian Gulf in the 1950's, \citet{wright2024depoliticizing} notes how pay and job type were divided along lines of race and nationality, with locals working in labor-focused roles, Indian migrants as tradespeople, and with Europeans and Americans filling positions of authority~\citep{wright2024unruly}. 

\textbf{Like how a discourse that it is an imperative to encompass every language can serve to justify labor conditions in the acquisition of data for models, a discursive tying of national ambition to oil production efforts served to motivate the imperative of oil production and of the involvement of local labor in these pursuits --- even characterizing advocacy against repressive labor conditions as acts against the state.} As \citet{ehsani2018disappearing} writes on Iranian government and oil companies: ``...[T]hroughout the twentieth century, it was the nationalist demand for the greater Iranianization of the workforce and especially of the management that was a recurring theme in negotiations with foreign oil companies, and not the modification of the repressive
labor practices. In fact, whenever labor activists or political organizers attempted to establish independent trade unions, foreign oil companies
and Iranian political elites united in treating them as a  subversive and
communist inspired threat to national security during the Cold War.'' Also tying labor to national goals, in 1953, Aramco broadcast the following as part of a speech in its Dhahran camp in Saudi Arabia: ``Let each and every one of you make it his primary goal to cherish truth and to work for the progress of the country. Whosoever of you shall leave the ranks or pose obstacles before or spread dissention among the people shall be dealt with by us in a way that will set him once more on the right path and will protect the nation against this evil influence''~\citep{vitalis2009ayyam}.\footnote{\citet{vitalis2009america} lays out how the role such rhetoric played in the path to the company's eventual nationalization.}  

\section{Recommendations}\label{sec:lessons}
Through parallels from the fight for sovereign oil for the age of AI sovereignty's commodification, we propose considerations surrounding the measurement of technology transfer and the tracing of  data as a resource.

\subsection{Measuring Technological Transfer} 
 One reason oil-producing countries were left dependent on foreign expertise even after countries nationalized oil supply and production was that technological uptake in the Middle East was largely limited to the operation of vendor-specific machinery. This made the question of technology transfer  a pertinent policy issue in the 1970's~\citep{zorn1983permanent}. Even 100 years into the commodification of oil, most of the activities in oil-producing regions centered the utilization and operation of oil technologies, while engineering, research and development were concentrated elsewhere~\citep{asghari2013technology}. Indeed, \citet{ittmann2025technology} notes that, even while British oil companies established training and educational programs, these programs were focused ``enhancing worker productivity rather than producing technical experts capable of running oil operation''.\footnote{A 1984 report from the U.S. Office of Technology Assessment similarly highlights that deep technological knowledge transfer in the petroleum industry to the Middle East was limited~\citep{OTA1984technology} and that there were anxieties surrounding how research and development activities were generally not being conducted locally.}

One way that technology transfer is  measured in the oil context is through the ``In-Kingdom Value Add Score'', presented by Aramco after the company's nationalization to measure the degree to which Aramco's suppliers were contributing to broader economic efforts in Saudi Arabia. The score is calculated through factors like the training of Saudi locals, research and development activity in the country, as well as salaries paid to Saudis~\citep{itasaudiaramco2020}. There may be analogous aspects of technological transfer and absorption that could help qualify AI sovereignty claims.  For example, components of NVIDIA's NeMo ecosystem are featured prominently throughout sovereign AI development steps in NVIDIA's Technical Guide to Building Sovereign AI Models~\citep{nvidia2024sovereign}, including discussion of NeMo's desirable technical features.\footnote{NeMo is accessed and supported through NVIDIA's AI Enterprise and described as `` a modular, enterprise-ready software suite for managing the AI agent lifecycle—building, deploying, and optimizing agentic systems—from data curation, model customization and evaluation, to deployment, orchestration, and continuous optimization''.} While NeMo is available through open-source, productionizing with these tools requires a license, whose pricing is tied to the number of the company's GPUs, a type of AI chip, being used.\footnote{``These licenses start at \$4500 per GPU per year or ~\$1 per GPU per hour in the cloud. Pricing is based on the number of GPUs, not the number of NIMs. Pricing is the same regardless of GPU size''~\citep{nvidianimdocs}} This pricing model can then leverage the company's market dominance in chips and GPUs to tether clients to a vendor-specific paradigm of AI development.

Drawing from measures of technology transfer and absorption, the appraisal of corporate claims surrounding technological leadership  could differentiate the uptake of vendor-specific AI tools from support in the development of or education surrounding vendor-agnostic technologies. Another factor to consider could include the difference between technological absorption versus consumption, and relevant questions here could include the level of access~\citep{solaiman2025beyond} institutions have into the sovereign models, such as how the models are hosted and  whether the mechanisms through which models are modified depend on
vendor pipelines.

\subsection{Tracing Data as a Resource}
Permanent sovereignty is a principle that has championed the transition from oil concession agreements to joint ventures. As ``a central issue in the debate over the legality of nationalizing foreign enterprises''~\citep{atsegbua1993principle}, it was used to contest the right of nationalized foreign enterprises to exploit another nation's national resources. The concept was first defined in the 1962 United Nations Resolution on Permanent Sovereignty over Natural Resources~\citep{tyagi2015permanent}, which declares that ``[t]he right of peoples and nations to permanent sovereignty
over their natural wealth and resources must be exercised in the interest of their national development and of the well-being of the people of the State concerned''~\citep{UNGA1962}. 

NVIDIA CEO Jensen Huang has motivated the importance of treating data as a kind of ``natural resource''~\citep{Mayfield2025JensenHuang}, that data should not be exported outside of those borders to be ``refined'' through the production of AI models and then sold back to the country~\citep{Mayfield2025JensenHuang}. This rhetoric hearkens to oil with his use of the term ``refine'' and how, like in other countries, oil produced in Iran was sold back to Iran ``at exorbitant prices''.\footnote{``Finally, the company was carrying away
the oil, refining it, and selling it back to Iranians at exorbitant prices.''~\citep{shafiee2018machineries}.} Yet, the purview of Huang's treatment is tied to a notion of territorial data sovereignty --- that the ownership of data is tied to where data is produced. So long as data is produced from a territory, there is sovereignty over it. However, rather than being actively recorded or generated by a country's people, data may instead be extracted \emph{through the use and development of AI by these communities}.  Two ways to operationalize this concern in  alignment with permanent sovereignty's emphasis on the ``interest of their national development and of the well-being of the people of the State concerned'' could be: (1) through an understanding of data governance as including autonomy over how energy is used to process data and to what ends, and (2) the tracing of data to their ultimate use cases. \citet{veale2023rights} discusses how the local processing of data, framed as a measure taken in light of data protection and privacy laws, and considers the autonomy of people to shape their  participation in large scale computations. There are whispers of this paradigm of thinking taking shape, with complaints alleging violation of Illinois' Biometric Information Privacy Act pointing to the use of battery energy in efforts to cognize the harms that come from the possession of biometric data~\citep{yew2022regulating}. Data could also be surfaced in contractual arrangements as an asset being generated, exchanged, even extracted through the continued use of AI sovereignty services. The inclusion of this consideration could enable the scrutiny of the extent to which data generated using a community's energy resources and labor are similarly used in the service of community interests. As we detail in Section~\ref{sec:cloud}, technical measures can be used by oil companies and CSPs to retain control over their operations. At times, however, the same technical measures can also be an important tool to increase transparency and to scrutinize control over data. In a lobbying document from Google regarding the EU AI Act, the company noted that the development and use of privacy-enhancing technologies (PETs), the same basis of technologies that comprise sovereignty controls, should be incentivized, and thus, the proposed dataset transparency requirements should be lessened~\citep{google2021consultation}. However, PETs can also play a role in increasing transparency surrounding the content of datasets--- with methods like differential privacy potentially enabling a privacy-preserving count of the kinds of content in a particular dataset.

\section{Conclusion}
The hope in this paper is that we encourage a new way of seeing AI sovereignty's discourse and commodification and that we spur further inquiry into how history can be generative for disentangling sovereignty in the AI age. Discourses of tech boosterism and specifically as they are tied to new technological developments can be ahistorical.\footnote{``For some particular reason, technology is considered ahistorical,
and it provokes thought not about the past, but about the present and the future''~\citep{rurup1974historians}.} This ahistoricism can serve to amplify the  discourse that commodifiers of AI sovereignty put forth, allowing tech companies to define and address issues on their own terms. Through a tracing of sovereignty’s history discursively and in oil production, we aim to interrogate the terms through and on which AI companies are selling sovereignty.  It is not lost on us that countries in the EU and the UK --- behind oil companies in our discussions --- are now some of the biggest proponents of sovereignty in the technological and AI context~\citep{uk_sovereign_ai_unit_2025, eurostack_2024, floridi2020fight}. Latent to our paper's argument may be another one --- that looking to the conduct of their governments in the context of oil and resource sovereignty, more broadly, may also be important in the disentanglement of what it means for the continent to achieve sovereignty in the AI age.

Moving forward, we are in the midst of greater shifts in how sovereignty discourse may be  wielded and commodified: AI companies are attempting to move their operations into space~\citep{lee2025starcloud, beals2025exploring}, potentially sidestepping questions of sovereignty as they have defined them along meanings of  territory, and the lines between companies and governments are ever blurred~\citep{primack2025trump, bond2025openai}. In the midst of changes to the value's commodification in the horizon, it is important to remain vigilant to who has the authority to assign AI sovereignty.\footnote{\citet{singh2025rethinking} notes that sovereignty as it is being wielded in the context of AI is ``increasingly framed as power to''.} Amidst the dizzying discursive assignments of sovereignty along the AI stack, it is important to keep \citet{azoulay2019potential}'s reminder in mind, that: ``sovereignty is not a gift''.
\bibliographystyle{ACM-Reference-Format}
\bibliography{aaai25}

\end{document}